\documentclass[]{pasj01}
\Received{2021 June 29}
\Accepted{2021 September 22}
\Published{$\langle$publication date$\rangle$}

\usepackage{mathpazo}
\usepackage[T1]{fontenc}
\usepackage{natbib}
\usepackage{makecell, multirow}
\usepackage{tabularx, booktabs}
\usepackage[dvipsnames]{xcolor}
\usepackage{dblfloatfix}

\graphicspath{ {./} }

\begin{document}


\title{Multi-frequency radio observations of the radio-loud magnetar XTE J1810-197}

\author{Sujin \textsc{Eie}\altaffilmark{1,2}}
\altaffiltext{1}{Department of Astronomy, Graduate School of Science, The University of Tokyo, 7-3-1 Hongo, Bunkyo-ku, Tokyo 113-0033, Japan}
\email{sujin.eie@grad.nao.ac.jp}
\author{Toshio \textsc{Terasawa}\altaffilmark{2}}
\author{Takuya \textsc{Akahori}\altaffilmark{2, 3}}
\author{Tomoaki \textsc{Oyama}\altaffilmark{2}}
\author{Tomoya \textsc{Hirota}\altaffilmark{2,4}}
\altaffiltext{2}{Mizusawa VLBI Observatory, National Astronomical Observatory of Japan, 2-21-1 Osawa, Mitaka, Tokyo 181-8588, Japan}
\altaffiltext{3}{SKA Organization, Jodrell Bank, Lower Withington, Macclesfield, SK11 9DL, UK}
\altaffiltext{4}{Department of Astronomical Science, The Graduate University for Advanced Studies (SOKENDAI), 2-21-1 Osawa, Mitaka, Tokyo 181-8588, Japan}
\author{Yoshinori \textsc{Yonekura}\altaffilmark{5}}
\altaffiltext{5}{Center for Astronomy, Ibaraki University, 2-1-1 Bunkyo, Mito, Ibaraki 310-8512, Japan}
\author{Teruaki \textsc{Enoto}\altaffilmark{6}}
\altaffiltext{6}{Extreme Natural Phenomena RIKEN Hakubi Research Team, RIKEN Cluster for Pioneering Research, 2-1 Hirosawa, Wako, Saitama 351-0198, Japan}
\author{Mamoru \textsc{Sekido}\altaffilmark{7}}
\author{Kazuhiro \textsc{Takefuji}\altaffilmark{7,8}}
\altaffiltext{7}{Kashima Space Technology Center, National Institute of Information and Communications Technology, Kashima, Ibaraki 314-8501, Japan}
\altaffiltext{8}{Japan Aerospace Exploration Agency, Usuda Deep Space Center, 1831-6 Oomagari, Kamiodagiri, Saku, Nagano 384-0306, Japan}
\author{Hiroaki \textsc{Misawa}\altaffilmark{9}}
\author{Fuminori \textsc{Tsuchiya}\altaffilmark{9}}
\altaffiltext{9}{Planetary Plasma and Atmospheric Research Center, Tohoku University, 6-3 Aoba, Aramaki, Aoba-ku, Sendai, Miyagi 980-8578, Japan}
\author{Shota \textsc{Kisaka}\altaffilmark{10, 11, 12}}
\altaffiltext{10}{Frontier Research Institute for Interdisciplinary Sciences, Tohoku University, Sendai 980-8578, Japan}
\altaffiltext{11}{Astronomical Institute, Tohoku University, Sendai 980-8578, Japan}
\altaffiltext{12}{Department of Physical Science, Hiroshima University, Higashi-Hiroshima 739-8526, Japan}
\author{Takahiro \textsc{Aoki}\altaffilmark{13}}
\altaffiltext{13}{The Research Institute for Time Studies, Yamaguchi University, 1677-1 Yoshida, Yamaguchi 753-8511, Japan}
\author{Mareki \textsc{Honma}\altaffilmark{1,2,4}}

\KeyWords{pulsars: general --- stars: magnetars --- pulsars: individual (XTE J1810-197)}

\maketitle


\begin{abstract}

We report on the multi-frequency multi-epoch radio observations of the magnetar, XTE J1810-197, which exhibited a radio outburst from December 2018 after its 10-year quiescent period. We performed quasi-simultaneous observations with VERA (22~GHz), Hitachi (6.9~GHz and 8.4~GHz), Kashima (2.3~GHz), and Iitate (0.3~GHz) radio telescopes located in Japan to trace the variability of the magnetar radio pulsations during the observing period from 13 December 2018 to 12 June 2019. The pulse width goes narrower as the observing frequency goes higher, analogous to the general profile narrowing behavior of ordinary pulsars. When assuming a simple power law in the range of 2.3~GHz and 8.7~GHz, the radio spectrum of the magnetar goes steeper with the average spectral index $ \langle \alpha \rangle \approx -0.85$ for the first four months. The wide-band radio spectra inferred from our observations and the literature suggest that XTE J1810-197 would have a double-peaked spectrum with a valley point in 22 -- 150~GHz, where the first spectral peak infers a gigahertz-peaked spectrum (GPS) feature with a peak at a few GHz. The GPS and the high-frequency peak have been identified in the spectra of other radio-loud magnetars, thus they may be intrinsic features that can give a new insight to understand various emission mechanisms and surrounding environments of radio magnetars. Our study emphasizes the importance of simultaneous long-term broad-band observations toward radio-loud magnetars to capture the puzzling spectral features and establish a link to other types of neutron stars.
\end{abstract}


\section{Introduction} \label{Sec:Intro}

Magnetars, which are a subset of pulsars with strong X-ray/$\gamma$-ray outbursts, have long rotation periods (\,$> 1$\,s) and large spin-down rates, inferring that they are young ($\lesssim 10^4$\,yrs) neutron stars with extremely strong magnetic fields (\,$\geq 10^{14}$\,G) \citep{ThompsonDuncan1993}. While radio pulsation is a common nature of ordinary pulsars, only 6 out of $\sim$ 30 confirmed magnetars have shown detectable radio pulses as of November 2020 (\citealt{McGillCatalog}\footnote{McGill Online Magnetar Catalog: \\ http://www.physics.mcgill.ca/\textasciitilde pulsar/magnetar/main.html}). The radio pulses from magnetars have some different properties from those of ordinary pulsars. 

One different property is the radio frequency spectrum; the spectral index, $\alpha$, defined with the total intensity $I$ and the frequency $\nu$ as $I\propto \nu^\alpha$, is close to zero (e.g. \citealt{Torne2015}) and a gradual spectral steepening has been seen in some cases (e.g., \citealt{Lazaridis2008}), while ordinary pulsars show much steeper spectra with the average spectral index $ \langle \alpha \rangle = -1.8 \pm 0.2 $ \citep{Maron2000}. The flatter spectra of radio-loud magnetars lead to relatively stronger radio emission at high frequencies. It allows us to explore the new type of neutron star emission in much wider radio band.

Another different property is the strong time-variability of flux density, spectrum, pulse shape, and polarization (e.g., \citealt{Camilo2008, Lazaridis2008, Levin2010}). The variation trend is poorly understood, nevertheless the fluctuations are known to be extreme both in the short- and long-terms. Even, their radio emissions are transient phenomena; the appearance is possibly associated with X-ray outbursts (e.g., \citealt{Anderson2012}). The flatter spectra and erratic variations of magnetar radio pulsations have been thought to be intrinsic to magnetars themselves, suggesting a possibility that there are some different radio-emission mechanisms in neutron star magnetospheres. 

The triggers for their extraordinary features are not yet figured out despite numerous hypotheses on the origins of magnetars. Meanwhile, \citet{Camilo2016} suggested that the spectrum of XTE J1810-197 became steeper down to $\alpha \approx -3$ before it entered the radio quiescent period. Such a steep spectrum at a late phase of outburst looks like, or even steeper than, that of ordinary pulsars, which can infer a possible similarity in emission mechanism except that the magnetar pulsations faded out by unknown reasons. Very recently, in addition, the discovery of the fifth radio-loud source showed that the magnetar Swift J1818.0-1607 has a steep spectrum with $\alpha \simeq -2.26$ \citep{Lower2020} even soon after the radio outburst emerged. It opened a new paradigm that a magnetar can also emit radio pulses akin to those from ordinary pulsars. Its spectrum later flattened \citep{Majid2020bATel}, implying the variety of magnetar spectral changes. Because such behavior is known in only one example to date, it is important to examine spectral indices and time variability for other magnetar radio outbursts.

XTE J1810-197 is known as the first radio-loud magnetar \citep{Halpern2005a} as well as the first transient magnetar detected with a significant increase of its X-ray luminosity \citep{Ibrahim2004}. The rotational period, $P = 5.541~\rm s$, and the spin-down rate, $\dot{P} = 2.83 \times 10^{-11} ~\rm s \;s^{-1}$ \citep{Pintore2016}, imply the strength of the dipole magnetic field, $B_{\rm d} = 3.2 \times 10^{19} \sqrt{P \dot{P}} \approx 1.3 \times 10^{14} \, \rm G$, and the characteristic age, $\tau_{\rm c}=P/2\dot{P} \approx 31000$\,yr, both of which are typical values for magnetars. Radio pulsations of XTE J1810-197 were detected in 2006 with the very same period of its X-ray pulsations \citep{Camilo2006}. After that, the radio flux density dramatically decreased. Then, it turned out to be in a quiescent state from November 2008 \citep{Camilo2016}. Hereafter, we call the radio-bright period in 2005 -- 2008 ``the previous outburst'' in this paper.

After the long radio-silent phase of about 10 years, intense radio pulses from XTE J1810-197 were again observed on MJD 58460 (2018-12-08) \citep{Lyne2018ATel}. \citet{Gotthelf2019} reported an earlier increase of its X-ray flux between MJDs 58442 and 58448, and confirmed X-ray outburst has preceded the radio burst. Many radio follow-up observations have reported the broad frequency range of its radio activity during this outburst, from 300~MHz \citep{Maan2019} to a few GHz \citep{Levin2019, Dai2019, Johnston2020}, to 32~GHz \citep{Pearlman2019} and even up to 150~GHz and 260~GHz \citep{Torne2020}.

These extensive studies of XTE J1810-197 at multiple radio frequencies during the second outburst, though, mostly cover time spans shorter than two months and are less overlapped in the observing periods. This could restrict the comprehensive understanding of highly variable magnetar emission in frequency and time domains. Simultaneous broad radio-band observations of radio magnetars in a wide time span that can capture from the very first moment of an outburst would enable to trace their long-term emission evolution and put together low and high frequency characteristics, providing important aspects to understand the behavior of one of the early phases of neutron stars.

In this paper, we analyze our radio data of $\sim\,$monthly observations of XTE J1810-197 during its second outburst and investigate the temporal and frequency evolutions of its pulsed emission for six months. This paper is organized as follows. In Section \ref{Sec:Obs}, we describe our observations and data reduction. We show the variations in spin parameters, pulse profile, flux, and spectrum in Section \ref{Sec:Results}. In Section~\ref{Sec:Discussion}, we discuss our results of the spectra and pulse widths and underline the significance of simultaneous long-term broad-band observations of radio magnetars.


\section{Observation and Data Reduction} \label{Sec:Obs}

We have conducted roughly-monthly, quasi-simultaneous, single-dish observations using radio telescopes in Japan. The observations started on MJD 58465 (2018-12-13), five days after the first report of the strong radio flares of XTE J1810-197 in late 2018 \citep{Lyne2018ATel}. Our observations were organized at 0.3~GHz, 2.3~GHz, 6.9~GHz, 8.4~GHz, and 22~GHz and they are summarized in Table~\ref{Table:Obs}. If the time difference between observations is shorter than $\sim$a day, we classified them in the same session assuming that averaged pulse shape would not change significantly within 24 hours.


\subsection{Observations}

\textit{VERA}: 
VLBI Exploration of Radio Astrometry (VERA) is a very long baseline interferometry (VLBI) array operated by the National Astronomical Observatory of Japan (NAOJ) \citep{VERA2020}. VERA consists of four 20-m diameter antennas located at Mizusawa, Iriki, Ishigaki, and Ogasawara in Japan. The director's discretionary time observations of XTE J1810-197 with VERA were conducted for four sessions, and the total amount of observing time per session is $\sim$ 4.2 hours including overhead. We note that VLBI was not achieved owing to a lack of a bright calibrator. In this study, therefore, we utilized the four VERA stations as separate four single-dish telescopes. Among them, Ishigaki station did not capture any pulsed signals which were observed at the other three VERA stations. It is likely due to bad weather at this station, therefore we exclude the data of Ishigaki station from our data reduction for this study. K-band (21.5 -- 23.5 GHz) receiver was used and the left-hand circular polarized (LCP) waves were recorded. The 22~GHz data of VERA was recorded after divided into four sub-bands (bandwidth of 512~MHz each) with the data recording rate of 2048~Mbps. Due to the weak intensity of the magnetar, we integrated the four sub-bands into one broad-band in order to increase the continuum sensitivity by a factor of 2. We then combined the three VERA stations into one dataset to earn an additional increase of the sensitivity (a factor of $\sqrt{3}$) after the phase is aligned by the barycentric dynamical time (TDB).

\textit{Hitachi}:
Hitachi 32-m telescope is located at Hitachi in Japan \citep{Yonekura2016}. Operated by Ibaraki University, Hitachi observations were performed for nine sessions. The C-band (6.6 -- 7.1 GHz) and X-band (8.1 -- 8.7 GHz) receivers recorded LCP and right-hand circular polarized (RCP) waves, respectively. Each band has the bandwidth of 512~MHz and the data recordings rate of 2048~Mbps.

\begin{figure*}[h!t!]
	\begin{center}
	\includegraphics[width=1.0\textwidth]{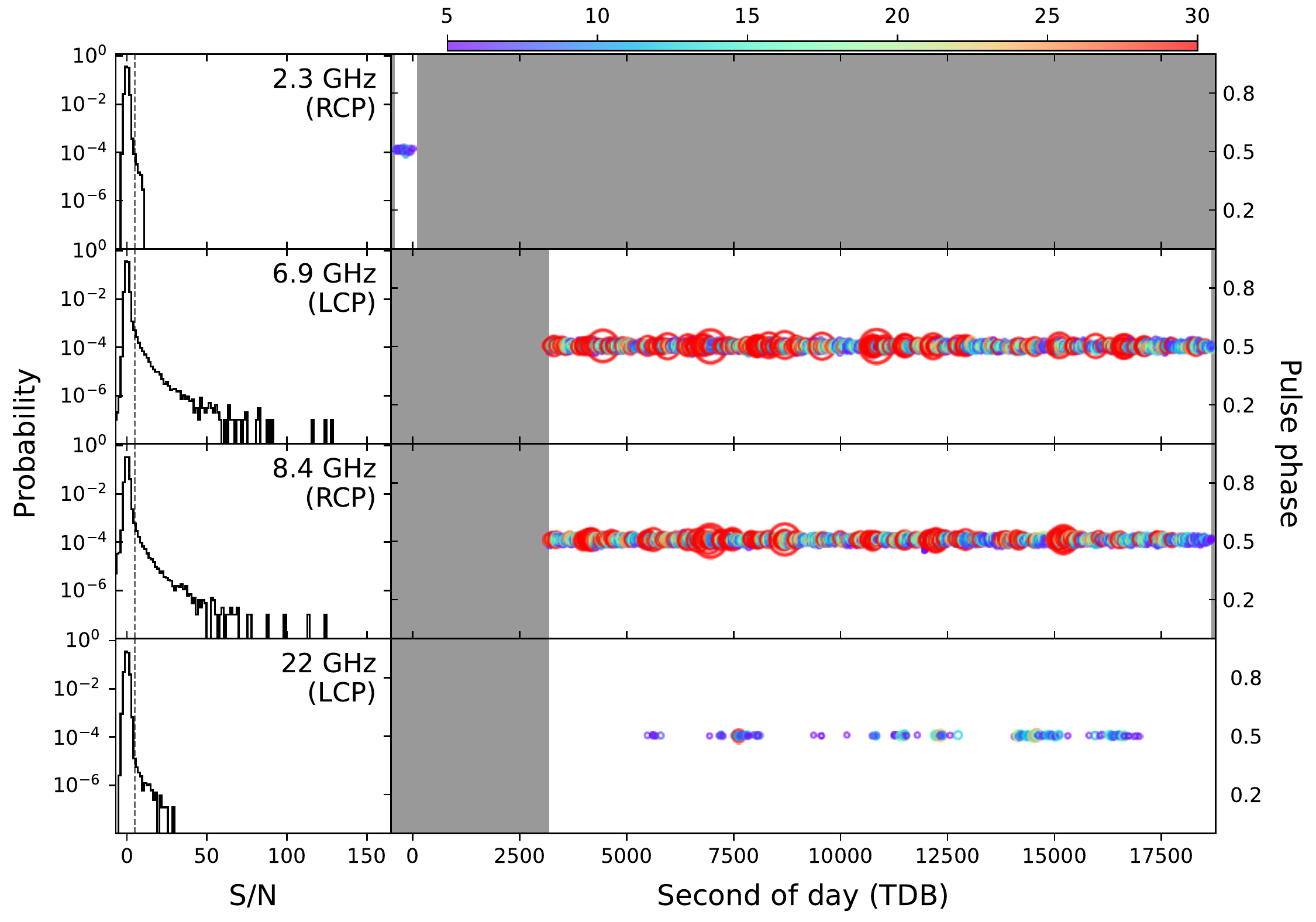}
	\end{center}
	\caption{
	\textit{Left panels:} The probability distributions of the S/N of single pulses on MJD 58490 (2019-01-07) after RFI cleaning. The vertical dashed line indicates the S/N value of 5, the plotting threshold for the right panels.
	\textit{Right panels:} Single pulses detected during our observations on the same date after RFI-cleaning in time-phase domain. 
	The optimal period and period derivatives on this observing epoch are deduced by the timing solution reported by \citet{Levin2019} (see Section~\ref{Subsec:Pevolution} for details), and the phase of the main peak is adjusted to be 0.5 based on the integrated profile at 6.9~GHz. Only signals with S/N $\rm>$ 5 are plotted and S/N values are indicated with the size and color scale shown by the color bar on the panels. 
	The gray-shaded areas indicate the time before and after the observations at each band. Kashima observation at 2.3~GHz was contaminated by strong RFIs $\sim$\,106\,s after the observation started.
	}
	\label{Fig:Detection}
\end{figure*}

\begin{table}[h!b]
\renewcommand{\arraystretch}{1.35}													
\setlength{\tabcolsep}{4 pt}													
\caption{Summary of observations. 
(1) Observing sessions. 
If observation dates are within a day for datasets, we consider them as the same session.
(2) UTC-based date and MJD.
(3) Center frequency and effective bandwidth.
(4) Polarization information: XY (Cartesian orthogonal), R (RCP), L (LCP)
(5) System equivalent flux density (SEFD).
(6) Start time of the observation in UT.
(7) Total observation time.
(8) Detection (Y) or non-detection (N) of the pulsed emission.
}
\label{Table:Obs}
\begin{center}
\footnotesize
\begin{tabularx}{\columnwidth}{@{\extracolsep{\fill}} c c l c c c c c @{\extracolsep{\fill}} }	

\specialrule{1pt}{0pt}{5pt}
	&	\makecell[c]{ Date \\[1pt] (MJD) }	&	\makecell[c]{ $\nu_{\rm center}$\,(BW) \\[1pt] {\scriptsize [MHz]} }	&	Pol.	&	\makecell[c]{ SEFD \\[1pt] {\scriptsize [Jy]} }	&	\makecell[c]{ $t_{\rm start}$ \\[1pt] { {\scriptsize [hh:mm]} }}	&	\makecell[c]{ $t_{\rm obs}$ \\[1pt] {\scriptsize [hr]} }	&		\\[2mm]
\makecell[l]{(1)}	&	\makecell[c]{(2)}	&	\makecell[c]{(3)}	&	\makecell[c]{(4)}	&	\makecell[c]{(5)}	&	\makecell[c]{(6)}	&	\makecell[c]{(7)}	&	\makecell[c]{(8)}	\\
\specialrule{.5pt}{1pt}{1pt}

1	&	2018-12-13	&	\,6856 ~~(512)	&	L	&	157.8	&	23:30	&	1.5	&	Y	\\
	&	(58465)	&	\,8448 ~~(512)	&	R	&	137.4	&	23:30	&	1.5	&	Y	\\[1mm]
	&	2018-12-14	&	\,325.1 ~(4)	&	XY	&	846.0	&	00:11	&	2.8	&	N	\\
	&	(58466)	&	\,2258 ~~(96)	&	R	&	532.0	&	00:10	&	4.0	&	Y	\\[1mm]
2	&	2018-12-18	&	\,6856 ~~(512)	&	L	&	139.1	&	02:00	&	4.0	&	Y	\\
	&	(58470)	&	\,8448 ~~(512)	&	R	&	122.3	&	02:00	&	4.0	&	Y	\\
	&		&	\,22483 (2048)	&	L	&	2944.4	&	02:44	&	4.2	&	Y	\\[1mm]
3	&	2019-01-07	&	\,325.1 ~(4)	&	XY	&	846.0	&	01:00	&	2.0	&	N	\\
	&	(58490)	&	\,2258 ~~(96)	&	R	&	532.0	&	00:00	&	0.5	&	Y	\\
	&		&	\,6856 ~~(512)	&	L	&	136.8	&	01:00	&	4.2	&	Y	\\
	&		&	\,8448 ~~(512)	&	R	&	123.8	&	01:00	&	4.2	&	Y	\\
	&		&	\,22483 (2048)	&	L	&	2820.5	&	01:24	&	4.2	&	Y	\\[1mm]
4	&	2019-01-09	&	\,6856 ~~(512)	&	L	&	137.8	&	22:40	&	0.6	&	Y	\\
	&	(58492)	&	\,8448 ~~(512)	&	R	&	129.0	&	22:40	&	1.3	&	Y	\\[1mm]
5	&	2019-01-21	&	\,22483 (2048)	&	L	&	2510.0	&	00:24	&	4.2	&	N	\\
	&	(58504)	&		&		&		&		&		&		\\[1mm]
6	&	2019-02-15	&	\,6856 ~~(512)	&	L	&	133.4	&	21:00	&	3.0	&	Y	\\
	&	(58529)	&	\,8448 ~~(512)	&	R	&	144.9	&	21:00	&	3.0	&	Y	\\
	&		&	\,22483 (2048)	&	L	&	4413.4	&	21:29	&	4.2	&	Y$^*$	\\[1mm]
7	&	2019-03-04	&	\,2258 ~~(96)	&	R	&	532.0	&	22:00	&	1.0	&	Y	\\
	&	(58546)	&	\,6856 ~~(512)	&	L	&	136.4	&	20:00	&	4.0	&	Y	\\
	&		&	\,8448 ~~(512)	&	R	&	129.2	&	19:00	&	5.0	&	Y	\\[1mm]
8	&	2019-03-31	&	\,6856 ~~(512)	&	L	&	133.2	&	18:00	&	4.0	&	Y	\\
	&	(58573)	&	\,8448 ~~(512)	&	R	&	131.1	&	18:00	&	4.0	&	Y	\\[1mm]
9	&	2019-04-23	&	\,6856 ~~(512)	&	L	&	139.2	&	16:00	&	4.0	&	Y	\\
	&	(58596)	&	\,8448 ~~(512)	&	R	&	131.1	&	16:00	&	4.0	&	Y	\\[1mm]
	&	2019-04-24	&	\,325.1 ~(4)	&	XY	&	846.0	&	16:00	&	3.6	&	N	\\
	&	(58597)	&	\,2258 ~~(96)	&	R	&	532.0	&	16:00	&	4.0	&	Y	\\[1mm]
10	&	2019-06-12	&	\,325.1 ~(4)	&	XY	&	846.0	&	13:00	&	2.0	&	N	\\
	&	(58646)	&	\,2258 ~~(96)	&	R	&	532.0	&	13:30	&	4.0	&	N	\\
	&		&	\,6856 ~~(512)	&	L	&	139.2	&	13:00	&	4.0	&	Y	\\
	&		&	\,8448 ~~(512)	&	R	&	131.1	&	13:00	&	4.0	&	Y	\\
\specialrule{1pt}{3pt}{-2pt}
\end{tabularx}
\end{center}
{\footnotesize{
$^*$ Only six single pulses were detected and an integrated profile was not obtained (see Section~\ref{Subsec:Flux} for details).}}
\end{table}	

\textit{Kashima}:
Kashima 34-m telescope had been operated by the National Institute of Information and Communications Technology (NICT) of Japan. The telescope was located at Kashima, Japan. We used the S-band (2210 -- 2350~MHz) receiver and recorded RCP waves. It is found that the second sub-band among four-divided frequency bands (bandwidth of 32~MHz each) of Kashima data is highly contaminated by radio frequency interferences (RFIs), hence 96~MHz in the recorded bandwidth was chosen for our analysis. 

\textit{IPRT}:
Iitate Planetary Radio Telescope (IPRT) is a dual symmetric offset parabolic antenna with an aperture size of 31-m $\times$ 16.5-m in 2~sets (aperture area of 1023~$\rm m^2$ in total; \citealt{Iwai2012}). The telescope is located at Iitate, Japan, and operated by Tohoku University, primarily for investigating the dynamic behavior of Jupiter's synchrotron radiation \citep{Tsuchiya2010} and solar radio emission at low frequencies. It has been used as well for the study of giant pulses from the Crab pulsar \citep{Mikami2016}. We conducted IPRT observations of XTE J1810-197 at P-band (323.1 -- 327.1~MHz) and recorded two orthogonal polarized waves.


\subsection{Data Reduction}

The data obtained at the frequency lower than 6.9~GHz were coherently de-dispersed \citep{Hankins1975, Book_Handbook} with the known dispersion measure (DM) of $178 \rm \ pc \ cm^{-3}$ {\citep{Camilo2006}} to correct the dispersion delay caused by free electrons in the interstellar medium. Later, each data was averaged in each 2-ms. The frequencies of 8.4~GHz and 22~GHz are high enough to be free from dispersion effect (see also \citealt{Torne2015}); with the known DM and the given bandwidth, the dispersion delays are 1.3~ms and 0.2~ms at 8.4~GHz and 22~GHz, respectively, which are smaller than the averaging time. Hence, we did not carry out a de-dispersion process to the data at the two frequencies and averaged data for each 2-ms.

Non-periodical RFIs were manually removed from the data in frequency and time domains to prevent any artifacts from being recognized as strong pulses. A general trend caused by atmospheric changes and/or receiver characteristics is as well examined and calibrated.
A pulsar-timing package \texttt{TEMPO2} \citep{Tempo2} is used for transforming the reference frame from each geographic location to the solar system barycenter, i.e., from the coordinated universal time (UTC) to TDB. 


\section{Analysis and Results} \label{Sec:Results}

We detected strong pulsed emission from XTE J1810-197 including a number of single pulses in most of our observing sessions in 2.3 -- 22~GHz bands. We could not detect any pulsed emission at 0.3~GHz over $5\,\sigma$. The $3\,\sigma$ upper limit on the mean flux density at 0.3~GHz is $\sim\,$13~mJy when assuming a duty cycle of 5\,\% (see Section~\ref{Subsec:Profile} for the assumed duty cycle). We show one sample of the dataset on MJD 58490 in Figure~\ref{Fig:Detection}. No presence of interpulse is observed during our observations (see also Dai et al. 2019), whereas during the previous outburst highly variable interpulse components were intermittently detected in 2006 at 0.72 pulse phase later than the main pulse phase, some of which were brighter than the main pulses \citep{Kramer2007, Lazaridis2008}.


\subsection{Spin parameters} \label{Subsec:Pevolution}

\begin{table*}[!b]
\renewcommand{\arraystretch}{1.45}
\setlength{\tabcolsep}{4 pt}
\caption{
Spin parameters of XTE J1810-197 estimated by \citet{Levin2019} and this work. Numbers in parenthesis represent $1\,\sigma$ errors in unit of the last significant digits.
}
\label{Table:Timing}
\begin{center}
\scriptsize
\begin{tabularx}{\textwidth}{@{\extracolsep{\fill}} c c c c l l l l c @{\extracolsep{\fill}} }
\toprule
Work	&	\makecell{Date range \\[2pt]{[MJD]}}	&	\makecell{Epoch \\[2pt]{[MJD]}}	&	\makecell[c]{$\nu_0$ \\[2pt]{[$\rm Hz$]}}	&	\makecell[c]{$\dot{\nu}_0$ \\[2pt]{[$\rm Hz \ s^{-1}$]}}	&	\makecell[c]{$\ddot{\nu}_0$ \\[2pt]{[$\rm Hz \ s^{-2}$]}	}	&	\makecell[c]{${\nu}^{\ldots}_0$ \\[2pt]{[$\rm Hz \ s^{-3}$]}}	&	\makecell[c]{${\nu}^{....}_0$ \\[2pt]{[$\rm Hz \ s^{-4}$]}}	& \makecell[c]{RMSE\\[2pt]{[period]}} \\[3pt]
\midrule
\makecell[l]{\citet{Levin2019}}	&	58460 -- 58507	&	58484.0	&	0.180 458 147(2)	&	$-2.575(4) \times 10^{-13}$	&	$-5.5(4) \times 10^{-20}$	&	$-1.7(2) \: \,\, \times 10^{-26}$	&	$4.2(10) \times 10^{-32}$ & 0.0007	\\
\makecell[l]{This work}	&	58465 -- 58597	&	58484.0	&	0.180 458 142(1)	&	$-2.52(1) ~~\times 10^{-13}$	&	$-4.1(2) \times 10^{-20}$	&	$~~~4.1(11) \times 10^{-27}$	&	\makecell[c]{-}	& 0.15 \\
\bottomrule
\end{tabularx}
\end{center}
\end{table*}

\begin{figure}[]
	\vspace{5pt}
	\begin{center}	
	\includegraphics[width=\columnwidth]{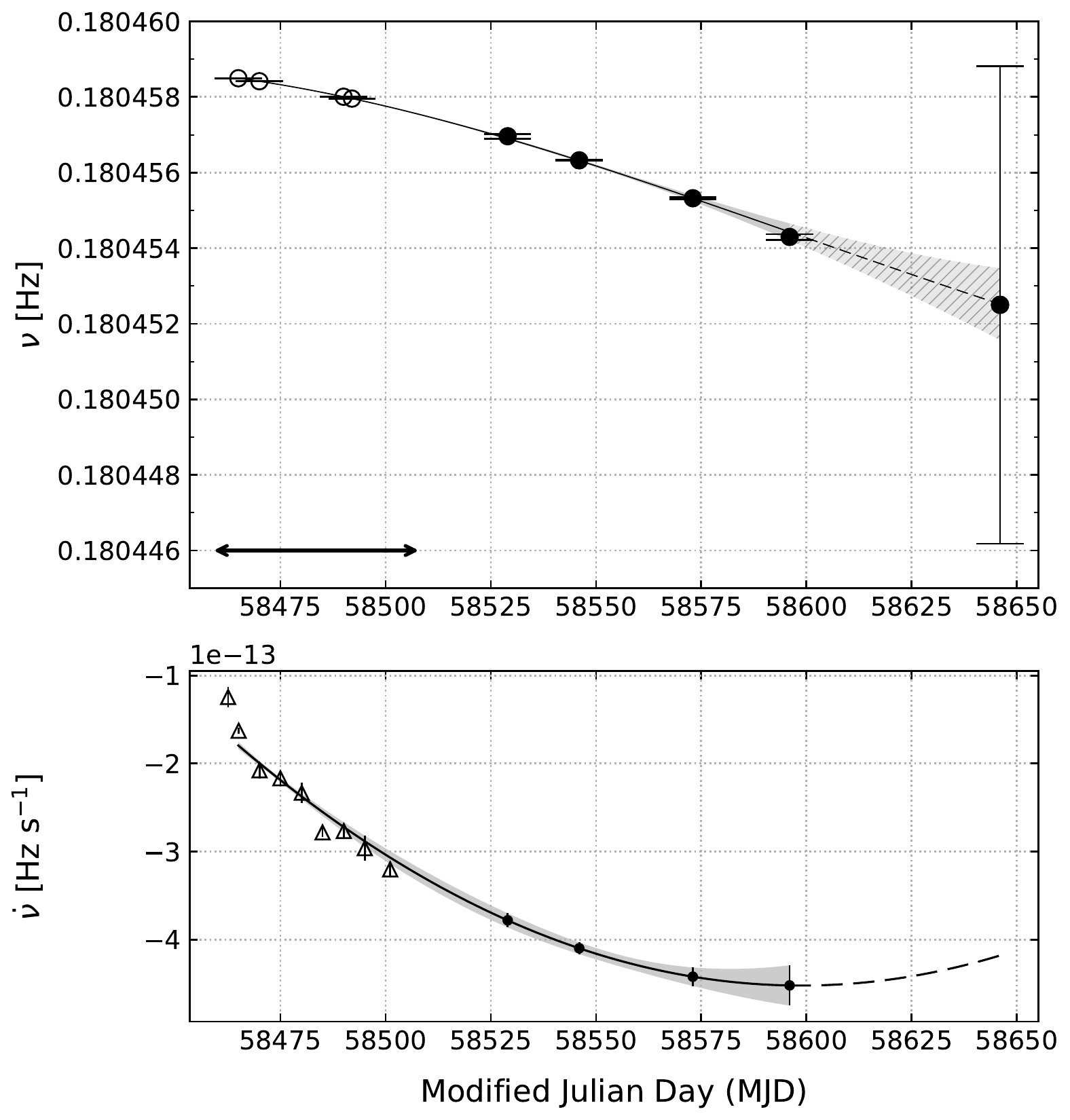}
	\end{center}
	\caption{
	\textit{Top panel:} Spin frequencies ($\nu$) of XTE J1810-197 during our observations. The arrow at the bottom left corner indicates the duration of 48 days when pulsar timing solutions were achieved by \citet{Levin2019}, i.e. MJDs 58470 -- 58507. Open circles are spin frequencies on our observing sessions 1 -- 4, which are derived by eq.~(1) with the spin parameters of \citet{Levin2019}. Filled circles are the values estimated from our data (see Section~\ref{Subsec:Pevolution} for the method). The curved line is a fitted variation of the rotation frequency (solid line: interpolation for MJDs 58507 -- 58596, dashed line: extrapolation of our fit to MJD 58646), and the shaded region encompasses $3\,\sigma$ confidence interval.
	\textit{Bottom panel:} Changes of the spin frequency derivative ($\dot{\nu}$) during our observing period. The solid line indicates $\dot{\nu}$ estimated from the upper panel, and the dashed line is an extrapolation of our fit to MJD 58646 which is excluded from the iteration process (see text for details). The shaded region encompasses $3\,\sigma$ confidence interval. The triangles are data taken from Fig. 6 in \citet{Levin2019}, while black dots indicate the estimated $\dot{\nu}$ on our observing sessions 6--9.
	}
	\label{Fig:Pevolution}
\end{figure}

\citet{Levin2019} determined spin parameters from their nearly-daily observations of XTE J1810-197 in the interval between MJDs 58460 and 58507, which includes our observing sessions~1\,--\,5. Single pulses obtained in these sessions are well aligned in phase with their timing results (Figure~\ref{Fig:Detection}). To follow up the timing parameters covering the whole observation period in this work, we determine spin parameters using arrival times of single pulse components at 8.4 GHz obtained by Hitachi telescope, except for session 5 when the Hitachi observation was not conducted. The determination process of spin frequency $\nu(t_i)$ (\textit{i} = 1\,--\,9) for session \textit{i} is in four steps. (\textit{\romannumeral 1}) We roughly find the optimal spin frequency, $\nu(t_i)$, that derives the maximum signal-to-noise ratio (S/N) in an integrated profile. (\textit{\romannumeral 2}) Grounded on the $\nu(t_i)$ estimated in the previous step, we search the frequency on closer inspection by conducting weighted least squares fitting to best align all single pulse components with $\rm S/N > 6$ at the same pulse phase throughout an observation, as is in Figure \ref{Fig:Detection}. It gives the consistent results with \citet{Levin2019} for the first four sessions. (\textit{\romannumeral 3}) We fit a Taylor polynomial (making use of polynomial package \texttt{NumPy)}
\begin{equation}
	\nu(t_i) = \nu_0 + \dot{\nu}_0(t_i - t_0) + \frac{1}{2}\ddot{\nu}_0(t_i - t_0)^2 + \frac{1}{6}{\nu^{\ldots}_0}(t_i - t_0)^3
\end{equation} \label{Eq:TaylorNu}
to $\nu(t_i)$ (\textit{i} = 1\,--\,9), and estimate derivatives of spin frequency for sessions 6\,--\,9. Here we remark that $\nu(t_i)$ (\textit{i} = 1\,--\,4; session 5 has non-detection) is obtained from Table~1 by \citet{Levin2019} which are derived from observations with a much higher cadence than ours. 
(\textit{\romannumeral 4}) Applying the derivatives, we perform step (\textit{\romannumeral 2}) again by aligning single pulses components in phase domain. Repeating steps (\textit{\romannumeral 3}) and (\textit{\romannumeral 4}) iteratively, we obtain $\nu(t_i)$ (\textit{i} = 6\,--\,9) and its derivatives. The last Hitachi observation on session 10 (MJD 58646; 2019-06-12) detected only a few weak single pulses at $>5\,\sigma$, therefore it is excluded from the above iteration process; i.e., only step (\textit{\romannumeral 1}) was conducted, resulting in a large uncertainty of $\nu(t_{10})$. The upper panel of Figure~\ref{Fig:Pevolution} shows the resultant $\nu(t_i)$ (\textit{i} = 6\,--\,10). The coefficients of the Taylor expansion~(Eq. 1) are listed in the second row of Table~\ref{Table:Timing}.

We confirmed the spin parameters obtained with the data at 8.4 GHz by applying them to align the single pulse components at 6.9 GHz which was simultaneously recorded with 8.4 GHz. Later we additionally verified that timing with \texttt{TEMPO2} gives consistent parameters with our method stated above.

By differentiating the Eq.~(1) with respect to $t_i$, we obtain the spin frequency derivative, $\dot{\nu}$, of the magnetar (Figure~\ref{Fig:Pevolution}, lower panel). Its absolute value $|\dot{\nu}|$ increased continuously, but the change became more gradual over time by MJD 58596, as was the case during the early 48-days as reported by \citet{Levin2019}. The similar trend in the X-ray band was also recently reported by \citet{Borghese2021}.

\begin{figure*}[htb!]
	\begin{center}
	\includegraphics[width=1.0\textwidth]{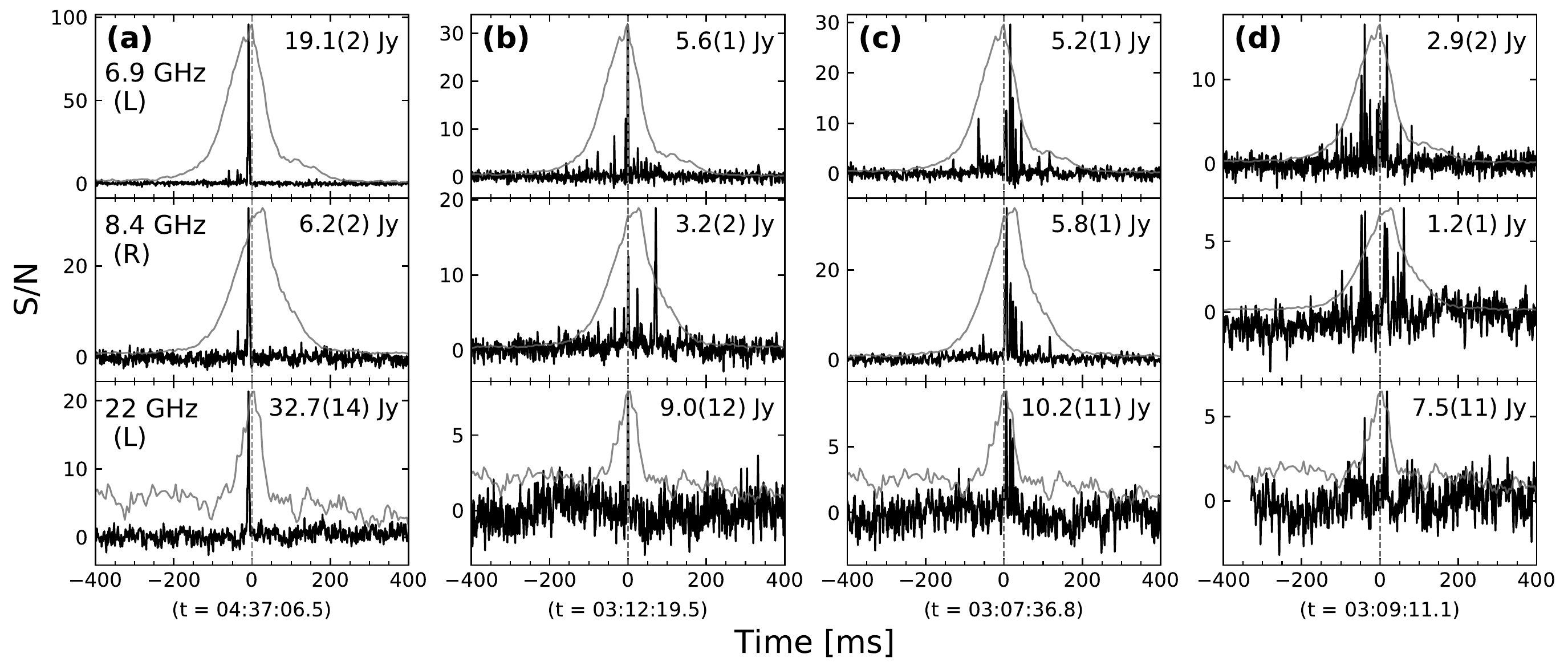}
	\end{center}
	\caption{
	Four examples of single pulses detected simultaneously at 6.9 (LCP), 8.4 (RCP), and 22~GHz (LCP) on MJD 58490 (2019-01-07). They are listed in order of complexity from left to right panels. The horizontal axes are aligned in TDB and set to zero at the peak phase of the average profile at 6.9~GHz, as is also in Figure~\ref{Fig:Detection}. The vertical dashed lines indicate the central phase of 0.5. The peak flux density of each single pulse profile is marked in Jy level, where its $1\,\sigma$ error in the last digit is shown in parentheses. The average profiles at the same frequency on MJD 58490 are shown as the solid gray lines, and scaled to the peak S/N in each panel.
	}
	\label{Fig:SPprofile}
\end{figure*}


\subsection{Pulse profiles} \label{Subsec:Profile}

Figure~\ref{Fig:SPprofile} shows some examples of single-pulse profiles (solid black lines) simultaneously observed at three different frequencies (6.9~GHz, 8.4~GHz, and 22 GHz). We note again that 6.9~GHz and 22~GHz indicate LCP waves, while 8.4~GHz does RCP waves. The shapes of the profiles are highly variable depending on time and polarization. Single pulses are approximately normally distributed within the pulse width, centered at the peak of the average profiles. We confirm that there is no periodicity in their occurrence in time and phase domains. We find that the profiles at 6.9~GHz and 8.4~GHz consist of more multiple components than those at 22~GHz. However, it is possible that the corresponding microstructures at a higher frequency have smaller relative strength so that are easily shrouded in strong noise. Some distinct examples of the multiplicity are shown in panels (c) and (d) of Figure~\ref{Fig:SPprofile}. In addition, strong pulse-to-pulse profile variations are prominently seen in Figure~\ref{Fig:SPprofile} when considering a direction of circular polarization. We could not differentiate that the variations are due to whether the change of the intensity of circular polarization or the real change of total intensity.

\begin{table*}[b!]
\renewcommand{\arraystretch}{1.25}
\setlength{\tabcolsep}{3 pt}
\caption{Profile widths at each direction of circular polarization (Pol.). For each observing frequency and MJD, pulse widths $W50_{\,\rm MJD}$ and $W10_{\, \rm MJD}$ are given in percentage of the duty cycle. The hyphen `--' indicates that an observation was conducted but the width is less than $3\,\sigma$ confidence.}
\label{Table:Pulsewidth}
\begin{center}
\scriptsize
\begin{tabularx}{\textwidth}{@{\extracolsep{\fill}} c c c c c c c c c c c @{\extracolsep{\fill}} }

\specialrule{1pt}{0pt}{4pt}
Pol.	&	\makecell[c]{ Observing \\[1pt] frequency}	&	\makecell[c]{ $W10_{58465}$ \\[1pt] {\scriptsize [\%]} }	&	\makecell[c]{ $W10_{58470}$ \\[1pt] {\scriptsize [\%]} }	&	\makecell[c]{ $W10_{58490}$ \\[1pt] {\scriptsize [\%]} }	&	\makecell[c]{ $W10_{58492}$ \\[1pt] {\scriptsize [\%]} }	&	\makecell[c]{ $W10_{58529}$ \\[1pt] {\scriptsize [\%]} }	&	\makecell[c]{ $W10_{58546}$ \\[1pt] {\scriptsize [\%]} }	&	\makecell[c]{ $W10_{58573}$ \\[1pt] {\scriptsize [\%]} }	&	\makecell[c]{ $W10_{58596}$ \\[1pt] {\scriptsize [\%]} }	&	\makecell[c]{ $W10_{58646}$ \\[1pt] {\scriptsize [\%]} }	\\[2mm]
\specialrule{0.5pt}{1pt}{1pt}
\specialrule{0.5pt}{0pt}{0.8pt}
\multirow{2}{*}{RCP}	&	2.3~GHz	&	5.7	&		&	--	&		&		&	--	&		&	--	&	--	\\
	&	8.4~GHz	&	3.8	&	4.3	&	4.9	&	4.6	&	--	&	4.6	&	3.2	&	--	&	--	\\
\specialrule{.5pt}{1pt}{1pt}																					
\multirow{2}{*}{LCP}	&	6.9~GHz	&	5.7	&	4.7	&	4.9	&	4.1	&	5.4	&	4.7	&	--	&	--	&	--	\\
	&	22~~GHz	&		&	4.2	&	2.4	&		&	--	&		&		&		&		\\
\specialrule{1pt}{1pt}{4pt}																					
	&		&	\makecell[c]{ $W50_{58465}$ \\[1pt] {\scriptsize [\%]} }	&	\makecell[c]{ $W50_{58470}$ \\[1pt] {\scriptsize [\%]} }	&	\makecell[c]{ $W50_{58490}$ \\[1pt] {\scriptsize [\%]} }	&	\makecell[c]{ $W50_{58492}$ \\[1pt] {\scriptsize [\%]} }	&	\makecell[c]{ $W50_{58529}$ \\[1pt] {\scriptsize [\%]} }	&	\makecell[c]{ $W50_{58546}$ \\[1pt] {\scriptsize [\%]} }	&	\makecell[c]{ $W50_{58573}$ \\[1pt] {\scriptsize [\%]} }	&	\makecell[c]{ $W50_{58596}$ \\[1pt] {\scriptsize [\%]} }	&	\makecell[c]{ $W50_{58646}$ \\[1pt] {\scriptsize [\%]} }	\\[2mm]
\specialrule{0.5pt}{1pt}{1pt}
\specialrule{0.5pt}{0pt}{0.8pt}									
\multirow{2}{*}{RCP}	&	2.3~GHz	&	3.0	&		&	2.4	&		&		&	2.5	&		&	1.1	&	--	\\
	&	8.4~GHz	&	1.8	&	1.6	&	2.1	&	2.2	&	2.1	&	2.3	&	2.0	&	2.5	&	2.7	\\
\specialrule{.5pt}{1pt}{1pt}																					
\multirow{2}{*}{LCP}	&	6.9~GHz	&	3.7	&	3.0	&	1.9	&	2.1	&	2.3	&	1.9	&	2.2	&	2.7	&		\\
	&	22~~GHz	&		&	1.4	&	0.9	&		&	--	&		&		&		&		\\
\specialrule{1pt}{1pt}{0pt}
\end{tabularx}
\end{center}
\end{table*}

The average profiles are overlaid in Figure~\ref{Fig:SPprofile} with solid gray lines. It is clear that single pulses have much narrower widths compared to the corresponding average profile. The pulse width is measured as the width at 10\,\% of the maximum intensity ($W_{10}$) of an integrated profile when the 10\,\% of the intensity is over $3\,\sigma$ (Table \ref{Table:Pulsewidth}). It derives a duty cycle of $\sim$5\,\% at 2.3 -- 8.4~GHz range which is comparable to the duty cycle during the previous outburst of XTE J1810-197 \citep{Camilo2016} as well as to the typical duty cycle of ordinary pulsars \citep{Maciesiak2011}. Since the 10\,\% of the maximum intensity of the profiles at 22~GHz are less than $3\,\sigma$, we measure the widths at 50\,\% of the peak intensity ($W_{50}$) and compare with $W_{50}$ at other frequencies in the same session. For each direction of the circular polarization, the pulse width $W_{50}$ (duty cycle likewise) goes slightly narrower as the observing frequency goes higher.

The average profile at 8.4~GHz is delayed when compared to those at 6.9~GHz and 22~GHz (see the phases at the peaks of solid gray profiles in Figure~\ref{Fig:SPprofile}). The degree of the phase delay is different at different observing epoch, but larger at earlier sessions. We presumed that the phase shift is caused by the time variations of the polarization degree, then we compared it with the polarization profiles of XTE J1810-197 in the literature. Taking into account a polarization study during MJDs 58463 -- 58470 at 0.7 -- 4~GHz \citep{Dai2019}, we derived LCP and RCP profiles from the total intensity ($I$) and circular polarization ($V$) profiles  by $I {\rm (LCP)} = (I+V)/2$ and $I {\rm (RCP)} = (I-V)/2$. The degree of the phase shift between our profiles at 6.9 and 22~GHz (LCP) and 8.4~GHz (RCP) on MJD 58470 (2018-12-18) is comparable with that between the two profiles at different circular polarization handedness taken from \citet{Dai2019} on MJD 58470. We, therefore, suggest the phase delay of averaged profiles inferred by Figure~\ref{Fig:SPprofile} would be on account of the different polarization degrees on each session.


\subsection{Flux variations} \label{Subsec:Flux}
To discuss the flux variations of XTE J1810-197, we estimate the total intensity, $I$, from single circular polarizations, LCP and RCP as follows. In the case of the data simultaneously taken with Hitachi (LCP at 6.9~GHz and RCP at 8.4~GHz), we simply combine them in each time bin (TDB) as $I$ = $I {\rm (LCP)} + I \rm (RCP)$ at $7.6\pm 1.1$~GHz. For the other data (Kashima 2.3 GHz and VERA 22 GHz), we measure conversion factors for converting the intensity at one circular handedness to total intensity. From the RCP and LCP profiles derived from the data obtained by \citet{Dai2019}, as described in Section~\ref{Subsec:Profile}, we calculate the ratio of mean flux densities of RCP or LCP profiles to the total flux density. Reciprocals of the ratios, $I$/LCP or $I$/RCP, are shown in Table~\ref{Table:Conversion}. We employ these conversion factors to derive the total intensities at 2.3~GHz and 22~GHz. Although the conversion factors can depend on the frequency, there is no significant change in the polarization profiles between 1920.0~MHz and 3712.0~MHz \citep{Dai2019}.
Therefore, we assume the polarization property at 3712.0~MHz would be stable up to 22~GHz, and apply the conversion factors obtained at 3712~MHz on MJD 58470 (2018-12-18) to our data at 22~GHz obtained on the same date. Additionally, the integrated profiles at 7.6~GHz do not appear to undergo significant temporal evolution during our observing period, we suppose that the averaged polarization pulse profiles and the averaged polarization angle are both relatively stable during this outburst. Therefore, we assume that the conversion factors do not significantly change during our observing period, and apply the mean conversion factors ($2.07 \pm 0.08$ and $2.01 \pm 0.09$ for LCP and RCP, respectively) to the other data except on MJD 58470.

\begin{table}[b]
\renewcommand{\arraystretch}{1}
\caption{Conversion factors from the intensity at single polarizations to the total intensity. The values are estimated from the profiles by \citet{Dai2019} (see text for details).}
\begin{center}
\footnotesize
\begin{tabularx}{0.95\columnwidth}{@{\extracolsep{\fill}} c || c c | c c | c c@{\extracolsep{\fill}} }
\specialrule{.8pt}{0pt}{0pt}
\multirow{3}{*}{\makecell{$\nu_{\rm center}$\\[3pt] {[MHz]}}}	&	\multicolumn{2}{c|}{MJD 58463}	&	\multicolumn{2}{c|}{MJD 58467}	&	\multicolumn{2}{c}{MJD 58470}
\rule[10pt]{0pt}{0ex}
\\
	&	\multicolumn{2}{c|}{\scriptsize{(2018-12-11)}}	&	\multicolumn{2}{c|}{\scriptsize{(2018-12-15)}}	&	\multicolumn{2}{c}{\scriptsize{(2018-12-18)}}
\\[.7mm]
	&	I/L	&	I/R		&	I/L	&	I/R	&	I/L	&	I/R
\rule[0pt]{0pt}{3pt}
	\\
\specialrule{.5pt}{0pt}{0pt}	
1024	&	1.97		&	2.03	&	2.29	&	1.77	&	2.06	&	1.94
\rule{0pt}{10pt}
\\[1mm]
1920	&	1.55		&	2.81	&	2.16		&	1.86	&	2.45	&	1.69
\rule[0pt]{0pt}{3pt}
\\[1mm]
2816	&	1.69	&	2.44	&	2.03	&	1.97	&	2.25	&	1.80
\rule[0pt]{0pt}{3pt}
\\[1mm]
3712	&	1.79		&	2.27	&	2.07	&	1.93	&	2.54	&	1.65
\rule[-.3pt]{0pt}{0ex}
\\
\specialrule{.8pt}{0pt}{0pt}
\end{tabularx}
\end{center}
\label{Table:Conversion}
\end{table}

\begin{figure}[ht]
	\begin{center}
	\includegraphics[width=1.0\columnwidth]{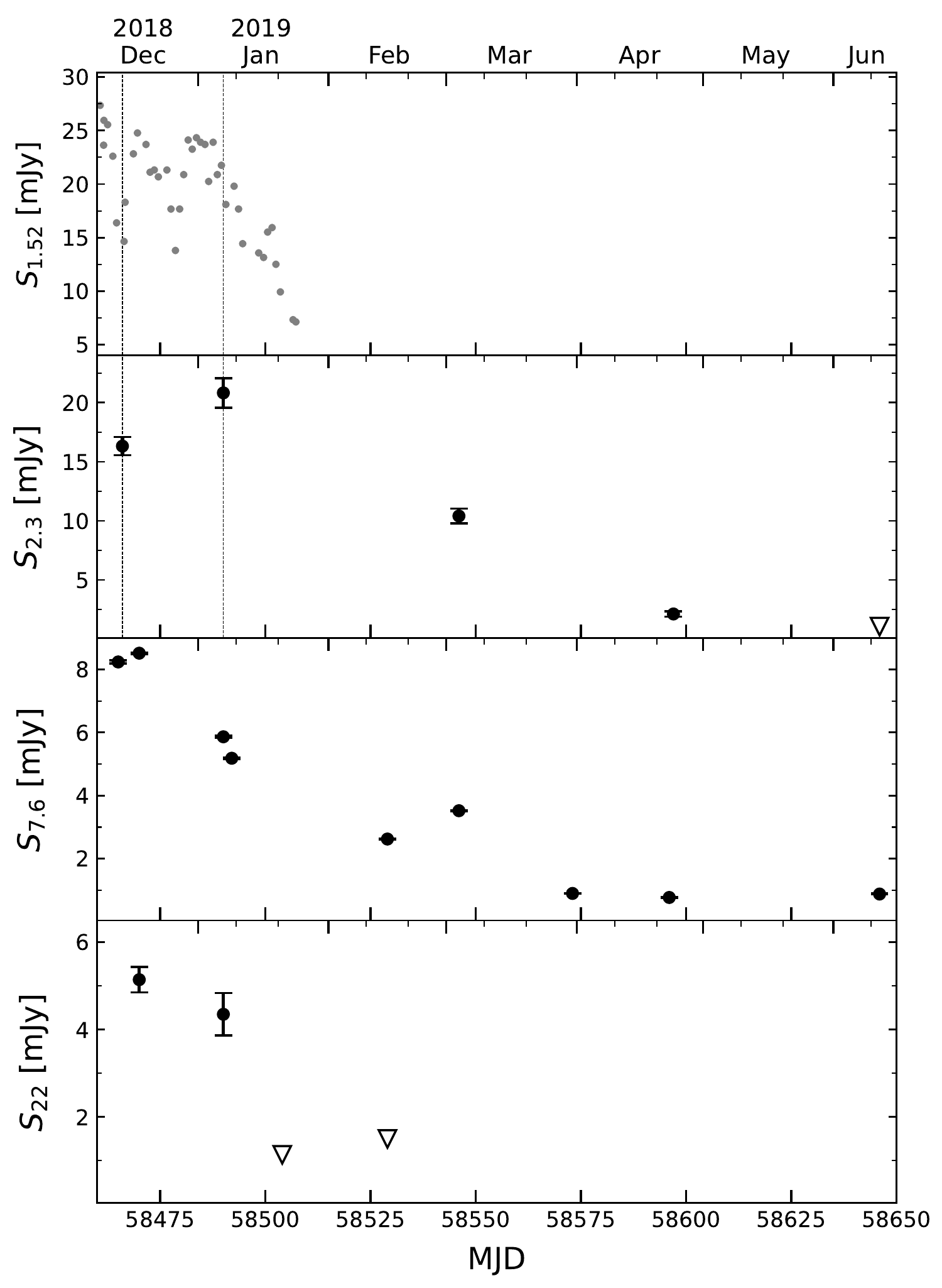}
	\end{center}
	\caption{
	\textit{Top panel:} Daily flux density variations at 1.52~GHz ($S_{1.52}$). Data are from \citet{Levin2019}.
	\textit{Lower panels:} Variations of total intensity of XTE J1810-197 at 2.3~GHz, 7.6~GHz, and 22 GHz ($S_{2.3}$, $S_{7.6}$ and $S_{22}$, respectively). Dashed vertical lines are for indicating the first session at 2.3~GHz is when the 1.52~GHz brightness has drastically decreased (first local minimum), whereas the second session is when the brightness is recovered from the next local minimum. Downwards pointing triangles indicate upper limits.
	}
	\label{Fig:Flux}
\end{figure}

As a result, Figure~\ref{Fig:Flux} shows time variations of total intensity at 2.3~GHz, 7.6~GHz, and 22~GHz. To compare and understand the variations with that at a lower frequency, daily flux densities at 1.52~GHz (data from \citealt{Levin2019}) are marked in the top panel of Figure~\ref{Fig:Flux}.

Mean flux density at 2.3~GHz was higher on MJD~58490 (2019-01-07) than that on MJD~58466 (2018-12-14). We note that it does not necessarily mean the intensity has gradually increased for a month; \citet{Levin2019} showed two drastic decreases in daily flux at 1.52~GHz within a month during this outburst and MJD~58466 is when the flux was in the first trough. Meanwhile, MJD~58490 is after the 1.52~GHz intensity recovered from the second trough (see vertical dashed lines in Figure~\ref{Fig:Flux}). Therefore, we remark that the early change at 2.3~GHz reflects erratic daily changes which is seen at the lower frequency. The last 2.3~GHz observation on MJD~58646 (2019-06-12) did not capture any pulsed emission and gives only an upper limit, as marked with a triangle in Figure~\ref{Fig:Flux}.

On the other hand, the flux density at 7.6~GHz does not seem to fully follow the low-frequency behavior on the same dates. On session 3 (MJD 58490; 2019-01-09), the intensity at 7.6~GHz is not brighter than the previous unlike the similar flux changes between 1.52~GHz and 2.3~GHz. The time-variable spectral indices are rather common for radio magnetars (e.g., \citealp{Anderson2012, Pennucci2015, Champion2020}), thus this would reflect a typical behavior of magnetar radio emission. Otherwise, it may suggest that there is a time lag in daily variations between low and high frequencies or that the flux variation includes narrow-band behaviors, which are not easy to be studied due to the strong variability.

Among four 22~GHz observations, we obtained integrated profiles only at the first and second sessions. We found the third session is when the luminosity at 1.52~GHz is two times fainter than at the second session \citep{Levin2019}, which may imply the general flux variations would be the wide-band feature. On MJD 58529 (2019-02-15) at 22~GHz, we found only six single pulses at $> 6\,\sigma$ significance due to the relatively low sensitivity. We thus marginally obtained the integrated pulse profile, hence no confident mean flux density is measured. The detection of a few reoccurred strong single pulses manifests the huge variability in flux density up to high frequency.

While the long-term variation of flux density might have been affected by daily variations, we see the radio intensity generally gets fainter over time. This result is in agreement with previous studies (e.g. \citealt{Camilo2007b}). During its previous radio outburst in 2005 -- 2008, \citet{Camilo2016} reported that XTE J1810-197 has shown a significant decrease in the mean flux density at 1.4 -- 3~GHz to a factor of 20 fainter than the first measured pulsed intensity during the first ten months, followed by a stable period for a twice longer period until it paused the radio emission. From our higher frequency observations, though, the intensity became $\sim$ ten times fainter at 7.6~GHz in six months after the outburst, therefore may not suggest a clear manifestation of radio cessation but infer a continuously decreasing phase in June 2019. Considering that the first radio detection in 2005 of XTE J1810-197 was a half year later than its X-ray outburst and an additional year has passed until the first detection of radio pulses, we estimate the time range of the observable decreasing phase in intensity would be longer for this outburst if the radio turn-off mechanism is analogous to its previous radio outburst. Further monitoring observations until this radio outburst come to the end, therefore, are needed to compare the radio behaviors and physical properties of XTE J1810-197 between the two outbursts.

\subsection{Spectral variations}

\begin{figure*}[ht]
	\begin{center}
	\includegraphics[width=\textwidth]{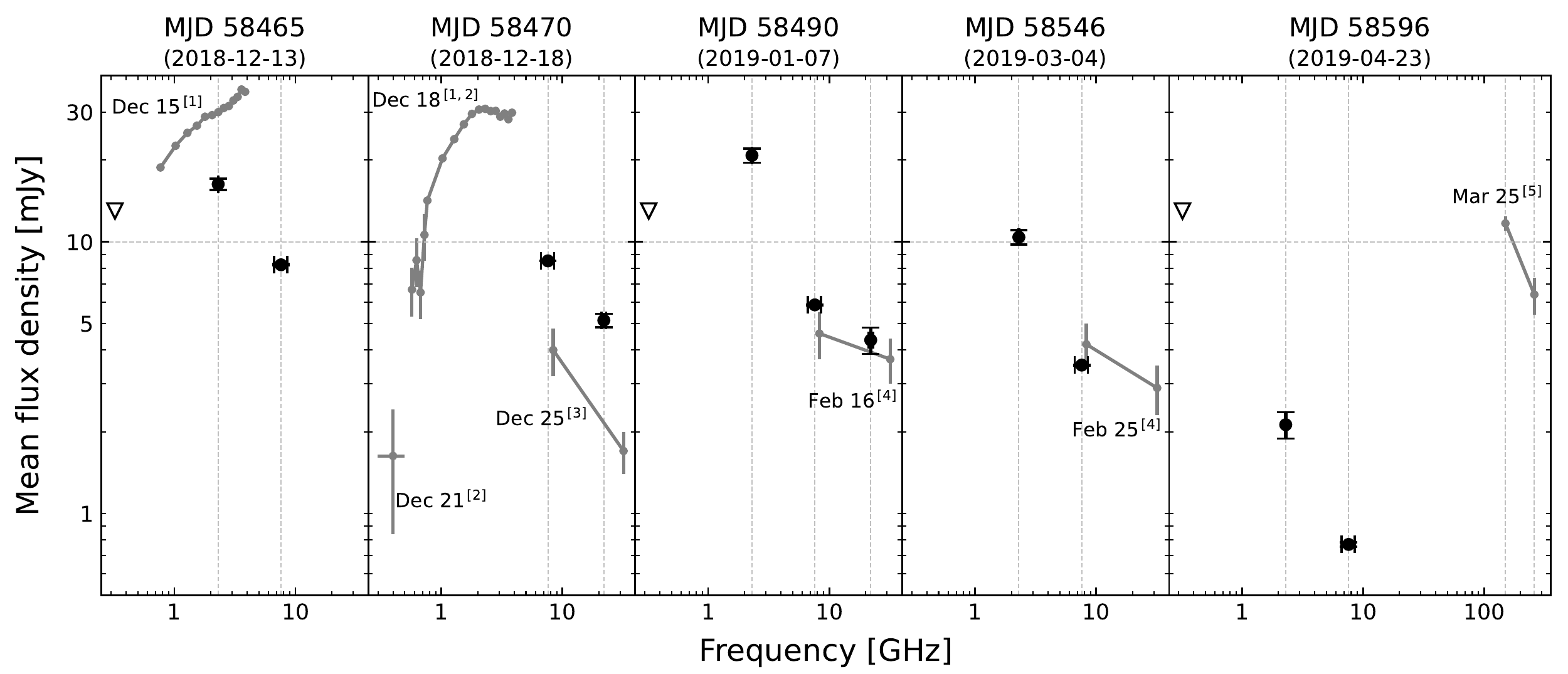}
	\end{center}
	\caption{
	Spectral variations of XTE J1810-197. Measured mean flux densities (converted to total intensity) at 2.3, 7.6, and 22~GHz are presented. For comparison, we added the flux densities obtained by [1] \citet{Dai2019}, [2] \citet{Maan2019}, [3] \citet{Pearlman2019}, [4] \citet{Pearlman2020arXiv}, and [5] \citet{Torne2020} with connected gray dots. Downwards pointing triangles indicate upper limits at 0.3~GHz.
	}
	\label{Fig:Spectrum}
\end{figure*}

We present the spectra of XTE J1810-197 between 2.3~GHz and 22~GHz obtained (quasi-) simultaneously. The time gaps between observations are smaller than 24~hours if not simultaneous. Figure~\ref{Fig:Spectrum} shows five spectra between December 2018 and April 2019, with early results of precedent studies (\citealt{Maan2019, Dai2019, Pearlman2019, Pearlman2020arXiv, Torne2020}). Our upper limit at 0.3~GHz is marked as well.

Even though each data point could be affected by short-term (shorter than a day) variations in flux density, our results suggest clear negative spectral indices in cm-wavelengths. The daily variations inferred by Figure~\ref{Fig:Flux} lead to the high variability in spectral indices, though we found the spectrum in 2.3 -- 8.7~GHz becomes steeper during the period when assuming a power law; $\alpha = -0.57 \pm 0.04$ on MJD 58465 to $-0.85 \pm 0.10$ on MJD 58596. During the previous outburst in 2006, the spectral index of the main pulse as well went steeper from $0.05 \pm 1.12$ (MJD 53873) to $-0.90 \pm 0.38$ (MJD 53983) in the frequency range of 1.4 -- 4.9~GHz during the first $\sim$4 months \citep{Camilo2007d}. Therefore, XTE J1810-197 shows a faster decline in its intensity toward higher frequency, both in the previous and this outburst.

The mean power-law spectral index in 2.3 -- 8.7~GHz during the 4.5 months of our observations is $\langle \alpha \rangle = - 0.85 \pm 0.14$. With regard to the spectrum up to 8~GHz, the previously measured index on MJD 53862 (2006-05-07) is $-0.30 \pm 0.04$ between 1.4 and 8.5~GHz \citep{Camilo2007d}, implying that the decreasing trends of the flux density at the radio frequency higher than 4~GHz are both observed in the previous and this radio outbursts.


\section{Discussion} \label{Sec:Discussion}

\subsection{Double-peaked spectrum?} \label{Subsec:double-peaked}
In Figure~\ref{Fig:Spectrum}, we overlay the flux densities obtained by \citealp{Dai2019} (0.7 -- 4~GHz), \citealp{Maan2019} (0.3 -- 0.7 GHz), \citealp{Pearlman2019} (8.4 and 32~GHz), \citealp{Pearlman2020arXiv} (8.3 and 31.9~GHz), and \citealp{Torne2020} (150 and 260~GHz). In contrast to that \citet{Maan2019} showed a monotonous positive spectral index at a broad low-frequency band of 0.3 -- 4~GHz by combining their results with the results of \citet{Dai2019}, the negative spectra at higher frequencies from our observations indicate that it is very difficult to explain the broad-band spectrum with a single power law, especially when we pay attention to the spectrum on MJD 58470 (2018-12-18).

In the frequency range up to $\sim$10~GHz, XTE J1810-197 shows a characteristic feature of gigahertz-peaked spectrum (GPS). To verify the GPS appearance even with the daily variations, we estimate a degree of flux variation from the daily observations at 1.52 GHz by \citet{Levin2019} (see the top panel of Figure~\ref{Fig:Flux}). The flux density has drastically decreased during the two local minima on MJDs 58466 and 58478; down to $\sim \! -35$\,\% in 2 days for both troughs. Even though assuming at most 50\,\% of daily flux changes randomly without a frequency-dependence, the flux density at 7.6~GHz is lower than the flux densities at 1.5 -- 3~GHz. Hence, XTE J1810-197 would still possess the GPS feature with a peak frequency between 1.5~GHz and 7.6~GHz. Moreover, the flux densities at adjacent frequencies would change similarly as inferred from Figure~\ref{Fig:Flux}, thus the overall shape of the spectra is expected to be relatively stable.

Such a GPS feature has already been discussed for radio pulsars and other radio-bright magnetars. Among known radio pulsars, eleven or more GPS pulsars have been identified \citep{Jankowski2018} and they may account for up to 10\,\% of the pulsar population \citep{Bates2013}. They typically have peak frequencies around 1~GHz or smaller \citep{Kijak2017}. Two radio magnetars, 1E 1547.1-5408 and PSR J1622-4950, have shown apparent GPS shapes with peak frequencies of 5~GHz and 8.3~GHz, respectively \citep{Kijak2013}. The turn-over features can be attributed to the thermal free-free absorption of the pulsar radio emission in peculiar environments, such as supernova remnants, pulsar wind nebulae, or dense H\,II regions \citep{Sieber1973, Kijak2011}. A radio magnetar located in the Galactic center, SGR J1745-2900, has been also suggested to be a candidate of GPS pulsar, while no apparent increase in flux density at low frequency was detected \citep{Lewandowski2015}. \citet{Lewandowski2015} suggest the free-free absorption in the electron material ejected during a magnetar outburst could be responsible for the GPS and the spectral evolution with peak frequency shifting toward lower frequency. As for XTE J1810-197, we could imply a long-term shift of the GPS peak to the lower frequency from two results; \textit{i}) the decrease of the peak frequency in 5-days in the early spectra at 0.7 -- 4 GHz \citep{Dai2019} and \textit{ii}) the steepening spectra at 2.3 -- 7.6 GHz from our observations. To further discuss the free-free absorption model by materials from the magnetar outburst to explain the spectral variations of XTE J1810-197, short-cadence observations around peak frequencies are needed for coming outbursts.

XTE J1810-197 has a more complicated spectral transition at higher radio frequencies over 20~GHz; the negative spectrum extending to 22~GHz is expected to face a spectral turn-up. Indeed, as marked in Figure~\ref{Fig:Spectrum}, \citet{Torne2020} has detected the magnetar emission on MJD 58567 (2019-03-25) at 150 and 260~GHz whose mean flux densities (11.7(7)~mJy and 6.4(10)~mJy, respectively) are higher than our upper limit at 22~GHz on MJD 58529 (2019-02-15; 2.6~mJy). The higher frequency intensities are even stronger than the intensity at 7.6~GHz obtained before and after their observation; the closest observation on MJD 58573 (2019-03-31) presents $0.90 \pm 0.01$ mJy at 7.6~GHz, suggesting that the spectrum would be inverted at over 22~GHz and has a second spectral peak between 22 -- 150 GHz. Here we remark the high-frequency peak is plausible even allowing daily flux fluctuations up to 50\,\%, as is confirmed for GPS.
Thus we suggest XTE J1810-197 would have a bimodal spectrum in the frequency range of 300~MHz to 250~GHz. 

Double-peaked spectral features are recently discussed for two other radio magnetars as well, SGR J1745-2900 and PSR J1622-4950, indicating that the emission mechanism could be different at low and high frequency ranges \citep{Chu2021}. However, including for XTE J1810-197, it is necessary to note that only simultaneous wide-band observations would confirm their extraordinary spectral shapes. But if true, they would be the first group with multiple spectral peaks in the radio band in the whole neutron star population. Although not all magnetar radio spectra have double-peaked features, this spectral hallmark could be intrinsic to radio magnetars, giving a new insight to study the radio emissions from neutron stars in wide frequency domain. For better constraints for magnetar emission models at different frequency ranges, tracing the timely variations of the double peaks would be highly required. Further multi-epoch broad-band observations toward radio-loud magnetars, therefore, would be necessary to investigate the origin and behavior of the double-peaked spectra.

\subsection{Pulse width narrowing behavior}
We showed the pulse width variations in Section~\ref{Subsec:Profile}. XTE J1810-197 during the second outburst shows the narrower average profile toward high frequency in the range of 2.3 -- 22~GHz. This inverse frequency-dependency of pulse width is one of the general features of plenty of ordinary pulsars \citep{Pilia2016}. A radius-to-frequency mapping (RFM; \citealt{Cordes1978}) is usually used to describe the narrowing of pulse profiles. The RFM model postulates that higher radio frequency emission originates from lower altitude of the neutron star, thus the pulse profile width would increase toward lower observing frequency. Most of the average profiles of XTE J1810-197 at 2.3 -- 22~GHz during our observing period have a single peak and we did not find a significant phase lag between pulse peaks at different frequencies, implying that the observed peak would be a core component when based on the RFM model. The model is, however, based on the dipole magnetic field structure and expects S-shaped polarization angle swings within the framework of rotating vector model (RVM). The dramatic changes of the polarization position angles of radio magnetars inhibit the classical RVM to be applied for magnetars (e.g. \citealp{Kramer2007}). Although the RFM model might not be plausible for the cases of radio magnetars, the profile narrowing trend of XTE J1810-197 could give a hint for any link between magnetar radio emission and standard pulsar emission in terms of magnetospheric structures. However, the pulse width variations of radio magnetars during an outburst, even only for XTE J1810-197, are yet poorly understood, hence high-sensitive long-term investigations on the variability in radio profiles of magnetars are required for each radio outburst.

A recent study of pulse widths for ordinary pulsars, though, estimated more than 20\,\% of ordinary pulsars show positive correlations between pulse width and observing frequency, implying the diversity in the energy spectrum distribution across the emission regions of neutron stars \citep{Chen2014}. While most radio magnetars have rather stable or slight narrowing pulse width over frequency (e.g. \citealp{Wharton2019, Lower2020}), PSR J1622-4950 showed not only profile narrowing from 1.4~GHz to 6.6~GHz, but slight broadening at 17~GHz \citep{Levin2012}. When it comes to radio-loud magnetars, only a few atypical properties are known to be commonly seen, hindering an integrated understanding of the radio-emitting magnetars altogether. Searching the additional common features of magnetar radio pulsations would enable a comprehensible understanding of them. We, therefore, stress the necessity of the investigations into the frequency dependence of pulse widths of radio magnetars through simultaneous multi-frequency observations, and this would enable a better understanding of the magnetospheric geometry of magnetars. We, in addition, note that in the long-term, radio magnetars can provide crucial samples for studying ultra wideband frequency dependency of pulse width with the benefit of their flat spectra.


\section{Summary and Conclusion}
We conducted an analysis of the simultaneous observations of the radio-loud magnetar XTE J1810-197 with radio telescopes in Japan at 0.3, 2.3, 6.9, 8.4, and 22~GHz. Our observations were organized for six months during the revived radio outburst detected in December 2018. Significant pulsed emissions were detected at the frequency range of 2.3 -- 22~GHz and the radio pulsations generally became fainter during our observations.

By employing the timing solutions for the early phase of this outburst from the literature, we found the change of the spin frequency derivative has continuously weakened since the radio outburst. We also report the duty cycle of the magnetar during this outburst is $\sim$\,5\,\%, which is analogous to its previous outburst and a number of ordinary pulsars. The pulse widths tend to become narrower as the observing frequency goes higher.

We obtained the negative spectra in 2.3 -- 22~GHz with the average spectral index of $-0.85 \pm 0.14$ at 2.3 -- 8.7~GHz. By combining with the flux densities at lower and higher frequencies from the literature, we suggest that XTE J1810-197 would have a bimodal spectrum with a GPS feature at lower 7.6~GHz and a second peak at over 22~GHz. Considering that the magnetar radio emission is highly variable, we put special emphasis on the importance of long-term simultaneous broad-band observations of magnetar radio outbursts to understand the origin and emission mechanisms of magnetar radio pulsations.


\section*{Acknowledgements}
We thank all staff of Mizusawa VLBI Observatory of NAOJ.
We are grateful to Katsunori Shibata for providing information for VERA data reduction.
Observations at the Hitachi 32-m telescope are supported by the Inter-university collaborative
project ``Japanese VLBI Network'' of NAOJ.
This work are supported by JSPS/MEXT KAKENHI grant numbers 17H01116 (M.H.), 18H01245 (S.K.), 18H01246 (S.K., T.T.), 19K14712, and 21H01078 (S.K.).

\bibliographystyle{apj}
\begin{bibliography}{./Sujin_reference_XTE1810}
\end{bibliography}

\end{document}